\documentstyle[12pt]{article}
\begin{document}
\begin{titlepage}
\begin{flushright}
TU-660\\
hep-th/0206222\\
Revised
\end{flushright}
\ \\
\ \\
\begin{center}
{\LARGE \bf
Holographic Charge Excitation \\
 on Horizontal Boundary \\
}
\end{center}
\ \\
\ \\
\begin{center}
\Large{
M.Hotta 
 }\\
{\it 
Department of Physics, Tohoku University,\\
Sendai 980-8578, Japan
}
\end{center}
\ \\
\ \\

\begin{abstract}
\ \\

We argue that states with nontrivial horizontal 
 charges of BTZ black hole  can be excited by ordinary falling matter 
including Hawking radiation. 
 The matter effect does not break the integrability condition of the
 charges on the horizon. Thus we are able to trace the proccesses in
which
 the matter imprints the information on the horizon 
 by use of the charged states. 
 It is naturally expected that 
 in the thermal equilibrium with the Hawking radiation  
 the black hole wanders ergodically 
 through different horizontal states 
 due to thermal fluctuation of incoming matter. 
 This fact strengthens plausibility of  
 the basic part of Carlip's idea \cite{c}. 
 We also discuss
 some aspects of the quantum horizontal symmetry 
 and conjecture how  the precise
  black hole entropy will be given from our point of view.

\end{abstract}

\end{titlepage}

\section{
Introduction
}

\ \\

The black hole entropy problem \cite{bh} has been attacked for long
years by many people
 and the effort yields  significant progresses. Especially, the state
counting on BPS branes has been achieved in the string theory and the
precise entropy form of the corresponding black branes, ${\cal A}/4G$,
is obtained, using the stability of the BPS states against the strong
coupling of the gravitational interaction \cite{sv}.  It is quite an
interesting result because 
 the analysis really proves that 
 the origin of the black hole entropy is entirely statistical. Even
though such a cruicial success has been performed, it should be 
 stressed that there still a lot of questions remain. For example, we do
not 
 know where the states contributing to the entropy 
 live in the macroscopical black-hole spacetimes.
 In addition, we must explain explicitly how the no-hair theorem  is 
 reconciled with the hairy black hole with large entropy.  
 In the string argument  the BPS states are counted just 
 in the weak coupling region, that is, in the flat spacetime. Hence no
real curved spacetime with horizon 
 appears in the step and 
 we cannot find any resolution in the argument to answer the above
questions.

 From the view point that the black hole horizon is essential 
to resolve the questions, Carlip \cite{c} 
proposed a quite stimulating 
scenario. He pointed out a possibility that a part of gauge freedom of 
the
general covariance becomes dynamical on the horizon and generates
physical states contributing to the entropy. The idea itself does not
depend on
 the spacetime dimensions and the gravity theories 
 (general relativity, dilatonic gravity and so on)  as 
 long as the theories possess the general covariance in the action.
 In his idea the reconcilation between 
 the no-hair theorem and the tremendous amount of the entropy 
 becomes fairy evident. The point is that 
 the charges mentioned in the no-hair theorem, 
 mass $M$, angular momentum $J$ and gauge charges $Q$,
 characterize just the geometrical property 
 of the black-hole spacetimes. 
 In fact the charges appear in geometrical quantities like curvature in
any 
  coordinate system. Hence the change of the charges means generation of 
  different geometry which cannot be yielded by any coordinate
transformation.
   On the other hand the horizontal charges  
  are proposed to be 
  generators of a part of general coordinate transformation. Therefore
 the transformation generated by the horizontal charges does not make 
  a different geometry at all. The situation may be easily understood by
  recalling that momentum $P$ of a black hole  is a 
  significant physical charge, but 
  translation generated by $P$ does not change the geometry itself.
 Clearly two quantum states 
 with different $P$ values are  orthogonal to each other, 
 thus the two states are physically different and, for example,  
 will contribute independently to 
 the partitional function and the entropy of a black-hole many-body 
  system. The similar mechanism will happen near horizon of a black hole 
  and the horizontal charges are expected
  to generate many different states 
  without deformation of the geometry, respecting the no-hair theorem.

 As a path to reach the universal realization of his basic idea 
  in arbitrary dimensions, 
 a lot of people including himself \cite{c1} tried to  
 find a Virasoro symmetry with classically nonvanishing
 central charge on the horizon in order to get the entropy via 
 Cardy's CFT formula \cite{v}. Unfortunately,
  various difficulties prevent the scenario from succeeding so
far\cite{av}.  
  For example, some models 
  need to pick up a circle $S^1$ without any convincing reason 
  in higher-dimensional geometries,
 some have ill-defined boundary charge  or 
  no classical central extension for Cardy's formula and 
 some suffer from ghost states with negative norm. 
 Hence fully-satisfactory resolution based on the Virasoro-algebra
scenario 
 has not been achieved yet to explain the entropy.

In order to reach the realization of essential part of the Carlip's
idea, 
 another way may be still open.  
It has  been pointed out \cite{hss} that the Schwarzschild black hole
admits  
 local time translation and angular 
 diffeomorphism on the horizon to be a well-defined 
asymptotic symmetry and that 
nontrivial representation with nonvanishing canonical charge 
 can be really constructed.  In four dimensions 
 the charges are labeled by the 
 spherical harmonic indices $(lm)$ like $Q_{lm}$. 
 The symmetry appears naturally 
 in arbitrary dimensions (higher than two dimensions) 
 and does not request any selection of a circle $S^1$ near the horizon. 
  Though the algebra does not have 
 any classical central extension, 
 the regular symmetry on the horizon is the first example, as far as we
know, 
 which possesses a nonsinglet representation 
  even in the classical theory. The number of the charges,
  which is classically infinite,
  looks so large as to incorporate huge degeneracy 
  of the black hole states into the representational space. 
Available values of the charges 
  distinguish a tremendous number of different physical states  and
  might generate the exact black hole entropy.

In this paper we prove that falling matter into the horizon does not
break
 the integrability condition of the charges in the pertubation level. 
Consequently it is argued that 
the states with the nontrivial charges of BTZ black hole are actually  
excited by ordinary falling matter 
including Hawking radiation. 
 From this observation it is naturally expected that, 
 in the thermal equilibrium with the Hawking radiation,  
 the black hole wanders ergodically 
 through different horizontal states 
 due to thermal fluctuation of incoming matter. 
 This fact strengthens plausibility of 
 the basic part of the Carlip's idea. 
 In the final section we discuss
 some aspects of the quantum symmetry and conjecture how  the precise
  black hole entropy will be given from our point of view.

\section{Horizontal Charge and Boundary Condition} 

\ \\

Let us consider the three-diemsional Einstein gravity with negative
cosmological constant. The action reads 
\begin{eqnarray}
S = \frac{1}{16\pi G}
\int d^3 x 
\sqrt{-g} (R+2\mu^2 ),
\end{eqnarray}
where the constant $\mu (>0)$ fixes the cosmological constant scale. 
In the system it is well known  that the black hole solution exists, so
called the BTZ
 black hole. The metric is given as follows.
 \begin{eqnarray}
ds^2 = -A(r) d{t'}^2 +\frac{dr^2}{A(r)}
+r^2 \left( d\phi ' - \frac{J}{2r^2}  dt' \right)^2 ,
\end{eqnarray}
\begin{eqnarray}
A(r) =\mu^2 r^2 -M +\frac{J^2}{4r^2}=
 \frac{\mu^2 }{r^2}
(r^2 -r^2_+ )(r^2 -r_-^2 ),
\end{eqnarray}
where 
$0\leq \phi' \leq 2\pi $, $r_+ \geq r_-$, 
$M$ and $J$ are  mass and anglular momentum of the black hole
 normalized in the planck unit and $r_+$ is the radius of the horizon.
Introducing a constant $\epsilon >0$, 
which will be set the planck scale later,
we consider the following regular transformation.
\begin{eqnarray}
&&
t=4\pi T_H \epsilon t', \\
&&
\rho = 4\pi T_H \epsilon 
\left[-\int \frac{dr}{A (r)} +\frac{1}{2\pi T_H} 
\ln (4\pi T_H \epsilon ) \right],
\\
&&
\phi =\phi' -2\pi T_H \epsilon\frac{ J}{r^2_+} t' ,
\end{eqnarray}
where $0 \leq \phi \leq 2\pi$ and
$T_H= \mu^2 \frac{r^2_+ -r^2_-}{2\pi r_+}$ is the Hawking temperature.
Using the transformation the BTZ metric is reexpressed  near horizon
$(\rho\sim\infty)$  as
\begin{eqnarray}
d\bar{s}^2
=
\Delta(-dt^2 +d\rho^2 ) +(r_+^2 +O(\Delta) ) d\phi^2 +O(\Delta^2).
\label{3}
\end{eqnarray}
Here  $\Delta$ has been  defined as
\begin{eqnarray}
\Delta =e^{-\frac{\rho}{\epsilon}} \sim 0.
\end{eqnarray}
Then we propose asymptotic metrices to the BTZ background in
eqn(\ref{3})
 in the same spirit of \cite{hss} as follows.
\begin{eqnarray}
\left[
\begin{array}{ccc}
g_{tt} & g_{t\rho} & g_{t\phi} \\
g_{\rho t} & g_{\rho \rho} & g_{\rho \phi} \\
g_{\phi t} & g_{\phi \rho} & g_{\phi \phi} 
\end{array}
\right]
=
\left[
\begin{array}{ccc}
-\Delta +O(\Delta^2) & O( \Delta^2 )  & O(\Delta) \\
O(\Delta^2)  & \Delta + O(\Delta^2) & O(\Delta ) \\
O(\Delta ) & O(\Delta ) & O(1) 
\end{array}
\right],\label{4}
\end{eqnarray}
where $O(\Delta^k )$ implies that
\begin{eqnarray}
\left| \lim_{\rho \rightarrow \infty}
\frac{O(\Delta^k )}{\Delta^k}
\right|
=|f (t,\phi)|<\infty .
\end{eqnarray}
Note that the boundary congruence defined by
\begin{eqnarray}
&&
\rho=\infty, \label{7}
\\
&&
\phi = const. \label{8}
\end{eqnarray}
is always null in regular spacetimes with 
the metric in eqn(\ref{4}). This implies that the boundary is just
 analogous concept to the Rindler horizon in the flat spacetime, 
 that is, any physical matter across the boundary 
 cannot come back  to the outside. 
 Taking account of the gravitational backreaction, 
 falling matter from the thermal bath 
 may temporarily shift the physical  position of the boundary 
 at $\rho =\infty$ apart from the global
 event horizon. (Here the shift should be measured by appropriate 
 geometrical length $d\tilde{l}$ between 
 the congruence and the horizon without gauge ambiguity.) 
 However the deviation is not expected to become so large 
 because the black hole swallows 
  both positive energy flux (associated with on-shell incoming matter)
and negative energy flux 
 (associated with quanta pair-created with positive-energy particles  
 of the Hawking radiation). The average magnitudes of the both fluxes 
 take the same value because of the thermal equilibrium. Hence the
average
 shift of the boundary position will be suppressed.

It is rather straightfoward to prove 
that infinitesimal coordinate transformations which preserve
 the asymptotic condition in eqn(\ref{4})  take the following form.
\begin{eqnarray}
&&
\xi^t = U(\phi) +O(\Delta ), \label{5}
\\
&&
\xi^\rho =O(\Delta ),
\\
&&
\xi^\phi = V (\phi) +O(\Delta ), \label{6}
\end{eqnarray}
where $U(\phi)$ and $V(\phi)$ are arbitrary 
periodic functions of $\phi$.  
The algebra of the generators is now easily written down. 
It is convenient to introduce mode expansions for $U$ and $V$:
\begin{eqnarray}
&&
U(\phi) =\sum_m U_m e^{im\phi},
\\
&&
V(\phi) =\sum_m V_m e^{im\phi}.
\end{eqnarray}
Then the generators are defined as
\begin{eqnarray}
&&{\cal Q}^{(T)}_m =e^{im\phi} \partial_t ,\\
&&{\cal Q}^{(\Phi)}_m =e^{im\phi} \partial_\phi ,
\end{eqnarray}
and the algebra is explicitly calculated as follows.
\begin{eqnarray}
&&
[ {\cal Q}^{(T)}_m ,\ {\cal Q}^{(T)}_{m'} ]\ =0, \label{400}
\\
&&
[{\cal Q}^{(\Phi)}_m,\ {\cal Q}^{(T)}_{m'} ]
=im' \hat{Q}^{(T)}_{m+m'},
\\
&&
[{\cal Q}^{(\Phi)}_m ,\ {\cal Q}^{(\Phi)}_{m'} ]
=i(m' -m)\hat{Q}^{(\Phi)}_{m+m'} .\label{410}
\end{eqnarray}
Stress that this is just a classical $U(1)$ current algebra.

It must be checked next 
that canonical charges of the transformations in eqns(\ref{5}) $\sim$
(\ref{6}) are well-defined or integrable. 
Decomposing the asymptotic metric into the ADM form as
\begin{eqnarray}
ds^2 = -N^2 dt^2 +h_{ij} (dx^i +N^i dt) (dx^j +N^j dt),
\end{eqnarray}
the canonical theory of the system can be constructed. Introducing 
the surface deformation vector:
\begin{eqnarray}
&&
\hat{\xi}^t =N\xi^t,
\\
&&
\hat{\xi}^i = \xi^i +N^i \xi^t,
\end{eqnarray}
the deviation of the canonical charge takes ordinary form as
\begin{eqnarray}
\delta Q_H [\xi]
&
=
&
\oint_{\Delta =0} d\phi
\left[
\frac{\sqrt{h}}{16\pi G}
\left( h^{ac}h^{b\rho} -h^{ab}h^{c\rho} \right)
\left(\hat{\xi}^t \delta h_{ab|c} -\hat{\xi}^t_{|c} \delta h_{ab}
\right)
\right.
\nonumber\\
&&\left.\ \ \ \ \ \ \ \ \  \ \ \ \  \ \ 
+2\hat{\xi}^a \delta \Pi^\rho_a -\hat{\xi}^\rho \Pi^{ab} \delta h_{ab}
\right].\label{9}
\end{eqnarray}
Using the asymptotic condition in eqn(\ref{4}), it is shown that
$\int \delta Q_H [\xi]$ really exists and is analytically integrable 
for the transformations in 
eqns(\ref{5}) $\sim$ (\ref{6}). 
This results in
\begin{eqnarray}
Q_H [\xi]
=\int^{2\pi}_0 d\phi
\left[
\frac{\sqrt{h_{\phi\phi}}}{16\pi G \epsilon} \xi^t
+2 \Pi^\rho_\phi \xi^\phi
\right]_{\rho =\infty}.\label{10}
\end{eqnarray}
The charges in eqn(\ref{10}) are divided into
 the following two parts. 
 The local time translation on the horizon is associated
  with the charges:
\begin{eqnarray}
Q^{(T)}_m = 
\int^{2\pi}_0 d\phi e^{im\phi}
\left[
\frac{\sqrt{h_{\phi\phi}}}{16\pi G \epsilon} 
\right]_{\rho =\infty}. \label{27}
\end{eqnarray}
The angular diffeomorphism on the horizon is associated with
the charges:
\begin{eqnarray}
Q^{(\Phi)}_m =
\int^{2\pi}_0 d\phi e^{im\phi}
\left[
2 \Pi^\rho_\phi 
\right]_{\rho =\infty}.
\end{eqnarray}

The finite transformation corresponding to the infinitesimal one in eqns
(\ref{5}) $\sim$ (\ref{6}) is given as 
\begin{eqnarray}
&&
t' = t +T(\phi) +O(\Delta),\label{11}
\\
&&
\rho' =\rho +O(\Delta),
\\
&&
\phi' = \Phi(\phi) +\Delta \Phi_1 (t, \phi ) +O(\Delta^2), \label{12}
\end{eqnarray}
where $T(\phi +2\pi ) =T(\phi)$,
$\Phi(\phi +2\pi) =\Phi(\phi) +2\pi$, $\Phi_1 (\phi +2\pi) =\Phi_1
(\phi)$
 and $d\Phi/d\phi >0$.
Let us consider a BTZ black hole metric with $J=0$ and $M=(\mu r_+)^2$:
\begin{eqnarray}
ds^2 = e^{-\rho'/\epsilon}(-{dt'}^2 +d{\rho'}^2 )+ r^2_+ d{\phi'}^2
+\cdots .
\end{eqnarray}
The asymptotic transformation induces the following
  excited  metric from the  metric.
\begin{eqnarray}
ds^2 &=&
\Delta (-dt^2 +d\rho^2 )
\nonumber\\
&&
-2\Delta [\dot{T}(\phi) -r^2_+ \dot{\Phi} (\phi) 
\partial_t \Phi_1 (t,\phi) ]dtd\phi
\nonumber\\
&&
+r^2_+ \dot{\Phi} (\phi)^2 d\phi^2 
-2\Delta \frac{r^2_+}{\epsilon^2}\dot{\Phi}(\phi) \Phi_1 (t,\phi) 
d\rho d\phi
+\cdots, \label{13}
\end{eqnarray}
where $\dot{T}=\partial_\phi T$ and $\dot{\Phi} =\partial_\phi \Phi$.
The charges of the excited metric in eqn(\ref{13})  are explicitly
evaluated 
as 
\begin{eqnarray}
&&
Q^{(T)}_m
=\frac{r^2_+}{16\pi G \epsilon}\int^{2\pi}_0  \frac{d\Phi}{d\phi}
e^{im\phi} d\phi, \label{50}
\\
&&
Q^{(\Phi)}_m
=
\frac{r^2_+}{16\pi G \epsilon}\int^{2\pi}_0  \frac{dT}{d\phi}
\frac{d\Phi}{d\phi}
e^{im\phi} d\phi .
\end{eqnarray}
Consequently it is clear that they certainly take nonzero values in
general, 
that is,
the representation is not singlet even in the classical theory.

In the next section we will discuss  falling matter effect.
So let us change the variables as follows.
\begin{eqnarray}
&&
u=t-\rho,
\\
&&
\rho =\rho, 
\\
&&
\phi =\phi.
\end{eqnarray}
The asymptotic condition in eqn(\ref{4}) is now rewritten as 
\begin{eqnarray}
\left[
\begin{array}{ccc}
g_{uu} & g_{u\rho} & g_{u\phi} \\
g_{\rho u} & g_{\rho \rho} & g_{\rho \phi} \\
g_{\phi u} & g_{\phi \rho} & g_{\phi \phi} 
\end{array}
\right]
=
\left[
\begin{array}{ccc}
-\Delta +O(\Delta^2) & -\Delta +O( \Delta^2 )  & O(\Delta) \\
-\Delta +O(\Delta^2)  & O(\Delta^2) & O(\Delta ) \\
O(\Delta ) & O(\Delta ) & O(1) 
\end{array}
\right],\label{16}
\end{eqnarray}
where $O(\Delta^k )$ means that
\begin{eqnarray}
\left| 
\lim_{\rho \rightarrow \infty}
\frac{O(\Delta^k )}{\Delta^k}
\right|
=
| F^{(k)} (u,\phi)  |<\infty .
\end{eqnarray}
For the condition in eqn(\ref{16}), 
the asymptotic symmetry is reexpressed as
\begin{eqnarray}
&&
\xi^u = U(\phi) +O(\Delta ), \label{20}
\\
&&
\xi^\rho =O(\Delta ),
\\
&&
\xi^\phi = V (\phi) +O(\Delta ).\label{21}
\end{eqnarray}
Because the ADM decomposition is not avaiable in the new coordinates,
we check the integrability of the symmetric charges via the covariant
phase-space formulation \cite{wald}. 
The deviation of the charge is now defined as
\begin{eqnarray}
\delta {\cal H} [\xi]
=\delta \int_{\partial C} \frac{1}{2} {\bf \epsilon}_{\beta\alpha\mu}
Q^{\beta\alpha}
+\int_{\partial C}{\bf \epsilon}_{\beta\alpha\mu} \xi^\beta
\Theta^\alpha,
\label{17}
\end{eqnarray}
where
\begin{eqnarray}
Q^{\beta\alpha}
=\frac{1}{16\pi G}
\left[ \nabla^\alpha \xi^\beta -\nabla^\beta \xi^\alpha \right]
\end{eqnarray}
and
\begin{eqnarray}
\Theta^\alpha
=\frac{1}{16\pi G}
\left[
g_{\mu\nu} \nabla^\alpha \delta g^{\mu\nu} -\nabla_\nu \delta
g^{\nu\alpha}
\right].
\end{eqnarray}
The first term of r.h.s in eqn(\ref{17}) is clearly integrable, however,
 integrability of the second term is not trivial. We  prove  
 using eqn(\ref{16}) 
  that the Wald-Zoupas integrability condition \cite{wz} strictly holds 
  for the transformations in eqns(\ref{20}) $\sim$ (\ref{21})
  and show that the second term in eqn(\ref{17}) is integrable.
 Actually the form of the local time translation charge, for example, 
 is given as
\begin{eqnarray}
Q^{(T)}_m =\frac{1}{16\pi G \epsilon}
\oint_{\rho =\infty} d\phi e^{im\phi}
\left[
1-2\epsilon\partial_u 
\right]
\sqrt{g_{\phi\phi}}. \label{23}
\end{eqnarray}
In the static case 
 the second term in eqn(\ref{23}) vanishes 
and eqn(\ref{23}) coincides precisely with eqn(\ref{27}) 
 in the canonical theory, as it should be.

\section{Excitations with the Horizontal Charge}

\ \\

In this section we argue that falling matter does not break the
asymptotic 
boundary condition in eqn(\ref{16}) at least in the perturbative level. 
 The horizontal states with nontrivial charges are actually 
 excited by the matter.

Let us consider the following spinless BTZ black hole solution 
with horizon radius $r_+$  as the background for the perturbation.
\begin{eqnarray}
d\bar{s}^2 
=-\frac{1}{\sinh^2 (\mu \rho )}
\left( du^2 +2dud\rho \right)
+r^2_+ \coth^2 (\mu\rho) d\phi^2 .\label{37}
\end{eqnarray}
In the background the action of free massive scalar field reads 
\begin{eqnarray}
S_{matter}
=
\int d^3 x\sqrt{-\bar{g}}
\left[
-\frac{1}{2}(\bar{\nabla} \varphi )^2
-\frac{M^2}{2} \varphi^2
\right],
\end{eqnarray}
where $M$ is the mass of the field. Solving 
the equation of motion of the field,
 the general in-coming-wave solution can be explicitly written down as
\begin{eqnarray}
\varphi
&=&\sum^\infty_{L=-\infty} \int^\infty_0 dE A_L^{in} (E)
\left[2 \sinh (\mu\rho)\right]^{-i\frac{E}{\mu}}
\left[\coth (\mu\rho)\right]^{-i\frac{L}{\mu r_+}}e^{iE(u+\rho) +iL
\phi} 
\nonumber\\
&&\ \ \ \ \ \ \ \ \times
F\left(a(E,L),b(E,L),c(E,L);-\sinh^{-2} (\mu\rho) \right)
\nonumber\\
&&+c.c,\label{34}
\end{eqnarray}
where
\begin{eqnarray}
&&
a=\frac{1}{2}\left(
1+\sqrt{1 +\frac{M^2}{\mu^2}}
\right)
+\frac{i}{2}
\left(
\frac{E}{\mu} -\frac{L}{\mu r_+}
\right),
\\
&&
b=\frac{1}{2}\left(
1-\sqrt{1 +\frac{M^2}{\mu^2}}
\right)
+\frac{i}{2}
\left(
\frac{E}{\mu} -\frac{L}{\mu r_+}
\right),
\\
&&
c=1+i\frac{E}{\mu}.
\end{eqnarray}
Near the horizon ($\rho \sim \infty$), the solution behaves as
\begin{eqnarray}
\varphi  =
\sum_L \int^\infty_0 dE 
\left(
A^{in}_L (E) e^{iEu +iL\phi}
+
A^{in}_L (E)^*  e^{-iEu-iL\phi}
\right) +O\left( e^{-\frac{\rho}{\mu}} \right).
\end{eqnarray}

Note that the quantum effect can be taken into account 
 simply by replacing the coefficients 
$A^{in}_L (E)$ and $A^{in}_L (E)^\ast$ with operators
${\hat{A}}^{in}_L (E)$ and ${\hat{A}}^{in}_L (E)^\dagger$
 which satisfy
\begin{eqnarray}
&&
[{\hat{A}}^{in}_L (E),\ {\hat{A}}^{in}_{L'} (E')]=0,
\\
&&
[{\hat{A}}^{in}_L (E),\ {\hat{A}}^{in}_{L'} (E')^\dagger ]=
\frac{1}{8\pi^2 r_+ E }\delta_{LL'} \delta (E-E' ).
\end{eqnarray}
In order  to solve the Einstein equation, let us introduce the
Chern-Simons
variables.  The dual variable of the spin connection $\omega_\mu^{ab}$
 is defined as
\begin{eqnarray}
\omega^c_\mu =\frac{1}{2}\epsilon^{cab} \omega_{\mu ab}.
\end{eqnarray}
Using $\omega^a_\mu$ and the triad variable $e^a_\mu$,
 let us define the following two Chern-Simons gauge fields.
\begin{eqnarray}
&&
A^a_\mu =\omega^a_\mu +\mu e^a_\mu ,
\\
&&
\bar{A}^a_\mu =\omega^a_\mu -\mu e^a_\mu .
\end{eqnarray}
Then the Einstein equation can be rewritten as
\begin{eqnarray}
&&
\epsilon^{\mu\nu\lambda}
\left[\partial_\nu A_{\lambda c}
-
\partial_\lambda A_{\nu c}
+
\epsilon_{abc} A^a_\nu A^b_\lambda 
\right] =16\pi G e T^\mu_c ,\label{30}
\\
&&
\epsilon^{\mu\nu\lambda}
\left[\partial_\nu \bar{A}_{\lambda c}
-
\partial_\lambda \bar{A}_{\nu c}
+
\epsilon_{abc} \bar{A}^a_\nu \bar{A}^b_\lambda 
\right] =16\pi G e T^\mu_c,\label{31}
\end{eqnarray}
where $T^\mu_c$ is the stress tensor of the matter.
Now taking variation of eqns(\ref{30}) and (\ref{31}),
we get the first-order perturbative equation of motion.
Here let us choose the gauge fixing as
\begin{eqnarray}
\delta A^c_u =\delta \bar{A}^c_u=0.\label{1021}
\end{eqnarray}
In the original variables $\omega$ and $e$, the conditions
 are rewritten as
\begin{eqnarray}
&&\delta e^a_u =0,\\
&&\delta \omega^{ab}_u =0.
\end{eqnarray}
Note that these six conditions fix both general covariance( with three
degrees of freedom) and local Lorentz symmetry (with other three degrees
of freedom).
Thus in principle they can be expressed only in terms of metric
variables $g_{\mu\nu}$. However the analytic achievement is too
complicated except a single condition $\delta g_{uu} =0$. Fortunately it
is found that we do not need detailed 
 expression of the gauge fixing in the metric variables.  
The perturbative equation corresponding to eqns (\ref{30}) and
(\ref{31}) 
 can be integrated out formally just 
 by using the expressions in eqn(\ref{1021}).
The result is summarized as follows.
\begin{eqnarray}
&&
\delta {\bf A}_\rho = \int^u_{-\infty} \exp[(u-u') K(\rho)] 
{\bf J}_\phi (u' , \rho ,\phi) du' ,
\\
&&
\delta {\bf A}_\phi = -\int^u_{-\infty} \exp[(u-u') K(\rho)] 
{\bf J}_\rho (u' , \rho ,\phi) du' ,
\\
&&
\delta {\bf \bar{A}}_\rho = 
\int^u_{-\infty} \exp[(u-u') \bar{K}(\rho)] 
{\bf J}_\phi (u' , \rho ,\phi) du' ,
\\
&&
\delta {\bf \bar{A}}_\phi = -
 \int^u_{-\infty} \exp[(u-u') \bar{K}(\rho)] 
{\bf J}_\rho (u' , \rho ,\phi) du',
\end{eqnarray}
where the matrices $K$ and $\bar{K}$ are defined as
\begin{eqnarray}
K =
\left[
\begin{array}{ccc}
0 & \beta & 0 \\
\beta & 0 & -\alpha \\
0     & \alpha & 0
\end{array}
\right],
\ \ 
\bar{K} =
\left[
\begin{array}{ccc}
0 & \beta & 0 \\
\beta & 0 & \alpha \\
0     & -\alpha & 0
\end{array}
\right],
\end{eqnarray}
and the functions $\alpha$ and $\beta$ are 
\begin{eqnarray}
&&
\alpha = \frac{\mu}{\sinh (\mu\rho)},
\\
&&
\beta =\mu \coth (\mu\rho).
\end{eqnarray}
The source terms of the matter is expressed as
\begin{eqnarray}
&&
{\bf J}_\rho =
8\pi G e \left[
\begin{array}{c}
\delta T^{\rho 0} \\
\delta T^{\rho 1} \\
\delta T^{\rho 2}
\end{array}
\right],
\\
&&
{\bf J}_\phi =
8\pi G e \left[
\begin{array}{c}
\delta T^{\phi 0} \\
\delta T^{\phi 1} \\
\delta T^{\phi 2}
\end{array}
\right],
\end{eqnarray}
where 
\begin{eqnarray}
\delta T_{\alpha\beta}
=
\nabla_\alpha \varphi \nabla_\beta \varphi
-\frac{1}{2}g_{\alpha\beta}\left[(\nabla \varphi)^2 -M^2 \varphi^2
\right].
\label{35}
\end{eqnarray}
Substituting the general solution in eqn(\ref{34}) into eqn(\ref{35})
and
 taking the limit of $\rho \rightarrow \infty$, it is  
 proven that
the asymptotic metric condition in eqn(\ref{16}) still holds because
 the perturbative correction induced by the matter field is given in
general as
\begin{eqnarray}
\left[
\begin{array}{ccc}
\delta g_{uu} & \delta g_{u\rho} & \delta g_{u\phi} \\
\delta g_{\rho u} & \delta g_{\rho \rho} & \delta g_{\rho \phi} \\
\delta g_{\phi u} & \delta g_{\phi \rho} & \delta g_{\phi \phi} 
\end{array}
\right]
=
\left[
\begin{array}{ccc}
 0 &  O( \Delta^2 )  & O(\Delta) \\
O(\Delta^2)  & O(\Delta^2) & O(\Delta ) \\
O(\Delta ) & O(\Delta ) & O(1) 
\end{array}
\right],\label{1022}
\end{eqnarray}
where the coordinate variables $(u,\rho)$ have been rescaled as
\begin{eqnarray}
&&
u' = 2\mu \epsilon u,
\\
&&
\rho' = 2\mu \epsilon \left( \rho +\frac{1}{\mu} \ln (2\mu \epsilon )
\right).
\end{eqnarray}
Here the first component in eqn(\ref{1022}), $\delta g_{uu} =0$, comes
just
from our gauge choice. Besides this there exist two other nontrivial 
constraints of our gauge fixing
among the metric components in eqn(\ref{1022}). However the point is
that 
whatever gauge constraints appears additionally in the metric 
the existence condition of the charge is satisfied  
as long as eqn(\ref{1022}) holds. Therefore 
we did not need to translate our gauge choice into the metric language.

Initially the horizontal charges $Q^{(T)}_m$ vanish 
 for the background metric in eqn(\ref{37}) except $m=0$.
 After the matter falls into the black hole, they 
 get the following corrections, which are generally nonzero. 
\begin{eqnarray}
\delta Q^{(T)}_m 
=
r_+ \oint d\phi
e^{im\phi}
\left[
\int^u_{-\infty}
\left(
\partial_u \varphi (u' ,\infty, \phi) 
\right)^2 du' 
\right].
\end{eqnarray}
Therefore it is concluded that 
the matter certainly excites the states 
with $Q^{(T)}_{m\neq 0}$ nonvanishing.

Here it may be worthwhile to stress that 
even if one may remove the charge at the late time 
 by an asymptotic transformation 
 in eqns(\ref{20}) $\sim$ (\ref{21}):
\begin{eqnarray}
\delta Q^{(T)'}_m (u =\infty) =0,
\end{eqnarray}
the initial nonzero deviations appear due to the transformation itself: 
\begin{eqnarray}
\delta Q^{(T)'}_m (u = -\infty ) \neq 0.
\end{eqnarray}
Hence not the absolute value but the difference of the charges 
 generated through the process  has physically significance.

Next, as a simple example, let us consider  a shock-wave solution 
of the matter field near the horizon:
\begin{eqnarray}
\left(
\partial_u \varphi 
\right)^2 
= \kappa (\phi) \delta (u) + O(\Delta),
\end{eqnarray}
where $\kappa(\phi)$ is arbitrary periodic positive function of $\phi$.
Passing across the boundary,
 the wave imprints the nonzero value on the charges $Q^{(T)}_m$ as 
\begin{eqnarray}
\delta Q^{(T)}_m (u>0)
=r_+ \oint e^{im\phi} \kappa (\phi) d\phi.
\end{eqnarray}
 Interestingly  note that
 if one observes the charges $\delta Q^{(T)}_m $ 
after the matter falling, the functional form of $\kappa(\phi)$
 is reproduced completely as follows.
\begin{eqnarray}
\kappa(\phi) = \frac{1}{2\pi r_+} \sum_m \delta Q^{(T)}_m e^{-im\phi}.
\end{eqnarray}
Thus the horizontal charge plays a role of recorder on the horizon.

For the quantum field the charges may be obtained by replacing the
classical 
stress tensor into expectation value of the stress operator in a vacuum
state:
\begin{eqnarray}
\delta Q^{(T)}_m 
=
r_+ \oint d\phi
e^{im\phi}
\left[
\int^u_{-\infty}
\left\langle 
\hat{T}_{--}(u' ,\infty, \phi) 
 \right\rangle du' 
\right],
\end{eqnarray}
where, due to the quantum effect, the flux 
$\left\langle \hat{T}_{--} \right\rangle$ 
can take negative values, losing the mass of the black hole. 
It is easy to see that 
the quantum raditation is also able to excite the horizontal states, 
 similarly to the classical field.

 If the black hole completely evapolates swallowing negative quantum
flux, 
 the information registered in the horizontal charges
 will be released into the space via the black hole radiation 
  because of the conservation law of the charges. (Note that 
 the black hole radiation does not always look isotropic 
 in the coordinate system in which 
 the asymptotic condition in eqn(\ref{16}) is satisfied at each time.)

\section{Discussion about Quantum Hair States}
\ \\

We analyzed so far the classical gravity theory and showed the existence
 of the nontrivial horizontal charges and its excited states. 
In this section  we try to make a hand-waving but quite suggestive  
 argument of the quantum gravitational states.

As shown in eqns(\ref{400}) $\sim$ (\ref{410}),
 the classical algebra is just U(1) current algebra in a circle.
Hence it  tempts us to suppose that
 its quantum version with central extension:
\begin{eqnarray}
&&
[ \hat{Q}^{(T)}_m ,\ \hat{Q}^{(T)}_{m'} ]\ =\frac{k}{2}m\delta_{m+m'},
\\
&&
[\hat{Q}^{(\Phi)}_m,\ \hat{Q}^{(T)}_{m'} ]
=-m' \hat{Q}^{(T)}_{m+m'},
\\
&&
[\hat{Q}^{(\Phi)}_m ,\ \hat{Q}^{(\Phi)}_{m'} ]
=(m -m')\hat{Q}^{(\Phi)}_{m+m'} +\frac{c}{12}(m^3 -m) \delta_{m+m' ,0}
\end{eqnarray}
 gives some inspiration of the quantum treatment of 
 the horizontal state counting. Though there may be some q-deformation
of
 the algebra in the real quantum system, let us investigate
  to what extent the above algebra leads us to the black hole entropy.
 
 Note that 
the reflection symmetry, $\phi\rightarrow -\phi$, prefers
representations 
 with $k=0$ and $c=0$ for the black hole system. 
If $k\neq 0$, define
\begin{eqnarray}
a_m = \sqrt{\frac{2}{|k|m}} Q^{(T)}_{ sgn(k)\times m}
\end{eqnarray}
 for positive $m$. Then the following relation holds.
\begin{eqnarray}
[a_m ,a^\dagger_{m'} ] =\delta_{m m'}.
\end{eqnarray}
In order to make unitary representation,   we must impose that
\begin{eqnarray}
a_m |0>=0
\end{eqnarray}
 for {\it only} positive $m$. Hence the prescription  prevents 
 the symmetry, $m\rightarrow  -m$, from being respected.
Similarly, if $c>0$, 
the vacuum state in the unitary representation  is defined as
\begin{eqnarray}
\hat{Q}^{(\Phi)}_m |0> =0
\end{eqnarray}
with $m \geq -1$. Therefore the reflectional symmetry is not realized in
the treatment. Thus the representation with $k=c=0$ seems favored from
the viewpoint of the angular reflection.
  
In the case with  $k=c=0$:
\begin{eqnarray}
&&
[ \hat{Q}^{(T)}_m ,\ \hat{Q}^{(T)}_{m'} ]\ =0, \label{40}
\\
&&
[\hat{Q}^{(\Phi)}_m,\ \hat{Q}^{(T)}_{m'} ] \label{41}
=-m' \hat{Q}^{(T)}_{m+m'},
\\
&&
[\hat{Q}^{(\Phi)}_m ,\ \hat{Q}^{(\Phi)}_{m'} ] \label{42}
=(m -m')\hat{Q}^{(\Phi)}_{m+m'} ,
\end{eqnarray} 
the structure of the algebra 
is quite similar to that of the Poincar\'e algebra, 
which is constructed by the momentum $P_a$ and  the Lorentz generator
$J_{ab}$.
 In fact, if  $\hat{Q}^{(T)}_m$ are regarded as $P_a$ and 
 $\hat{Q}^{(\Phi)}_m$ as $J_{ab}$, the skeleton structure of algebras
 in eqns (\ref{40}) $\sim$ (\ref{42})
 reads
\begin{eqnarray}
&&
[P,\  P] =0,
\\
&&
[J,\ P] = P,
\\
&&
[J,\ J] = J.
\end{eqnarray}
This precisely coinsides with that of the Poincar\'e algebra.
Consequently
 the irreducible unitary representation 
  may be constructed  by use of the Wigner's little-group argument
\cite{wigner}.
  
  Let us think a vector made of the charges of 
  the background in eqn(\ref{3}),   
$(\bar{Q}_0^{(T)} , \bar{Q}_1^{(T)} ,\bar{Q}_{-1}^{(T)} ,\cdots )$
with
\begin{eqnarray}
&&
\bar{Q}^{(T)}_m 
= \frac{r^2_+}{8G\epsilon} \delta_{m 0},
\end{eqnarray}
 as a reference vector for the little group of our symmetry. 
Then the little group is defined as a subgroup generated by 
a part of $\hat{Q}^{(\Phi)}_m$. Under the little-group transformation 
 the reference vector $(\bar{Q}^{(T)}_m)$ must remain unchanged: 
\begin{eqnarray}
\delta_L \bar{Q}^{(T)}_m =0.
\end{eqnarray}
After some easy manipulation, it is proven that the little group
 is generated only by $\hat{Q}^{(\Phi)}_0$. 
Thus the little group is the rigid rotation of the horizon and just has
O(2)
 group structure .

 Because the operators $\hat{Q}^{(T)}_m$  commute with each other,
  we can diagonalize simultaneously all the operators.
  In order to construct unitary representation, 
  we assume that $\hat{Q}^{(T)}_m$  are hermitian. 
 Along the Wiger's argument, let a state  $|{\bf 0} ;s>$ denote an
eigenstate of the operator
 $\hat{Q}^{(T)}_m$ with its eigenvalue $\bar{Q}^{(T)}_m $:
\begin{eqnarray}
\hat{Q}^{(T)}_m |{\bf 0} ;s> =\bar{Q}^{(T)}_m |{\bf 0} ;s>.
\end{eqnarray}
Here the additional index $s$ is prepared for possible 
intrinsic spin freedom which is transformed by the little group O(2). 

Next we define  the act of the little group on the states.
Because the little group does not change  the reference vector
 $(\bar{Q}^{(T)}_m)$, 
 the subspace spanned by $\{ |{\bf 0},s > \}$ must be a representation
 of the little group: 
\begin{eqnarray}
\hat{Q}^{(\Phi)}_0 |{\bf 0},s > = \sum_{s'} J_{s s'} |{\bf 0},s' >,
\end{eqnarray}
where  $[J_{ss'}]$ is arbitrary matrix representation of O(2) group.
 If one impose irreducibility on the representation, 
 it must be singlet, that is, $[J_{ss'}]$ is a real number which
 $s$ denotes:
\begin{eqnarray}
\hat{Q}^{(\Phi)}_0 |{\bf 0},s > = s |{\bf 0},s >. 
\end{eqnarray}
Note that in this algebra, 
subtraction of constant from $\hat{Q}^{(\Phi)}_0$ is always 
 admitted. Consequently 
  we can set $s=0$ without loss of generality. Therefore  
  the index $s$ will be suppressed later.

By virtue of eqn(\ref{41}),  the finite transformation operator of the
 anglular diffeomorphism $\hat{L}(\Lambda)$ is introduced as
\begin{eqnarray}
\hat{L}(\Lambda)^{-1} \hat{Q}^{(T)}_m \hat{L}(\Lambda)
=\sum_{m'} \Lambda_{mm'}\hat{Q}^{(T)}_{m'} , \label{60}
\end{eqnarray}
where $\Lambda$ is arbitrary element of 
the fundamental representation of the group and,   
 in the infinitesimal case: 
$\Lambda_{mm'} =\delta_{mm'} +m\varepsilon_{m' -m}$, 
 the operator $\hat{L}(\Lambda)$is reduced into the infinitesimal one:
\begin{eqnarray}
\hat{L}(\Lambda ) 
=1+\sum_m \varepsilon_m \hat{Q}^{(\Phi)}_m .
\end{eqnarray}
Using eqn(\ref{60}), it is shown that 
the state $\hat{L} (\Lambda)  |{\bf 0}>$ is an  
 eignestate with a new eigenvalue of $\hat{Q}^{(T)}_m $:
\begin{eqnarray}
\hat{Q}^{(T)}_m \hat{L} (\Lambda)  |{\bf 0}>
=\sum_{m'} \Lambda_{mm'} \bar{Q}^{(T)}_{m'}
\hat{L} (\Lambda)  |{\bf 0}>.
\end{eqnarray}
Let us define here 
coefficients $f_m$ as
\begin{eqnarray}
f_m =\frac{8G\epsilon}{r^2_+}
\sum_{m'} \Lambda_{mm'} 
\bar{Q}^{(T)}_{m'}.
\end{eqnarray}
The coefficients $f_m$ are related with the original 
function $\Phi(\phi)$
 of the eigenvalues in eqn (\ref{50}) as follows.
\begin{eqnarray}
\frac{d\Phi}{d\phi} =\sum_{m} f_{m} e^{-im \phi }. 
\end{eqnarray}
Note that $f_0=1$ for arbitrary eigenstates of $\hat{Q}^{(T)}_0$. 
Just as in the Poincar\'e algebra, we can consistently 
introduce positive norm
for properly rescaled eigenvectors:
\begin{eqnarray}
&&
|f_m> \propto \hat{L} (\Lambda)  |{\bf 0}>,
\\
&&
\hat{Q}^{(T)}_m
|f_m >
=
\frac{r^2_+}{ 8G \epsilon}f_m 
|f_m>,
\end{eqnarray}
that is, the following relations can be set up.
\begin{eqnarray}
<f_m | f'_{m'} > &=&
\prod_{m>0} \delta({\it Re}f_m -{\it Re}f'_{m'}) 
\nonumber\\
&&\times\prod_{m>0} \delta({\it Im}f_m -{\it Im}f'_{m'}) .
\end{eqnarray}
Gathering all possible eigenstates generated by $\hat{L}(\Lambda)$, 
 we make a vector space spanned  by  $\{|f_m > \}$. By  construction 
 it is clearly a unitary irreducible representation of the algebra.

Now let us 
try to discuss the number of the horizontal states roughly. 
Because the eigenstates $|f_m >$ 
 with different eigenvalues are orthogonal to each other,
  they must be treated as physically different states. 
Thus we should know the number of possible  values  of $f_m$ at each
$m$, 
  denoted by $N_m$.
Note that
the maginitude of the eigenvalues $|f_m|$ must not be larger than one: 
\begin{eqnarray}
|f_m | \leq 1.
\end{eqnarray}
This is easliy noticed from the following manipulation: 
\begin{eqnarray}
&&
1=\frac{1}{2\pi}
\int^{2\pi}_0 \frac{d\Phi}{d\phi} \left| e^{im\phi} \right|d\phi
=\frac{1}{2\pi}
\int^{2\pi}_0 \left| \frac{d\Phi}{d\phi} e^{im\phi} \right|d\phi
\nonumber\\
&&
\geq \left|\frac{1}{2\pi}
\int^{2\pi}_0 \frac{d\Phi}{d\phi} e^{im\phi} d\phi \right|
=|f_m|,
\end{eqnarray}
where we have used  $\oint d\Phi =2\pi$ and monotonically 
increasingness of $\Phi(\phi)$.

Though  
the eigenvalues run only in the bounded regions $|f_m| \leq 1$ as seen
above,
 they are essentially continuous and shows a band structure 
  in the representation.  The continuous spectrum makes  
  the state counting  confusing. As in the free gas system, the number
of
  states might be naively defined by the form of 
  $V^{(T)} d^2 f_m /(2\pi \hbar)^2$ 
  where $V^{(T)}$ is ``volume'' of the space canonically conjugate to
the
  $f_m$ space. However the principle of 
  determination of the value of $V^{(T)}$ has not been established and
seems 
  rather ambiguous so far. 
  
 Fortunately, the bounded regions are independent of $m$. 
  Therefore it may be sufficient that 
 we just assign a proper integer N to the number of the values 
 of each  $f_m$.
Therefore the number of the states of the black hole $N_{BH}$ 
 may be roughly expressed as
\begin{eqnarray}
N_{BH} \sim \prod_{m=1} N_m \sim  \prod_{m=1} N. \label{63}
\end{eqnarray}
However, it is noticed soon that the estimation is divergent because the
subscript $m$  can 
take any large integer. This means that the quantum
 treatment discussed here  is  too naive  to count the
 black hole states precisely and needs some (q-)modification. 
  The estimation in eqn(\ref{63}) seems to suggest that 
 the fully quantized gravity theory supplies some cutoff $m_{max} $, 
 analogously to that of noncommutative spheres, and improves the
estimation as
\begin{eqnarray}
N_{BH} \sim  \prod^{m_{max}}_{m=1} N = N^{m_{max}}.
\end{eqnarray}
Consequently the entropy will be  written down as
\begin{eqnarray}
S_{BH} =\ln N_{BH} \sim  \ln N^{m_{max}} .
\end{eqnarray}
Because the  cutoff $m_{max}$ is 
naturally expected of order of the horizon circumference  
divided by the Planck length $\epsilon_{pl}$:
\begin{eqnarray}
m_{max} \sim \frac{2\pi r_+ }{ \epsilon_{pl}},
\end{eqnarray}
by taking the gravitational constant as
\begin{eqnarray}
G = \frac{\epsilon_{pl}}{4 \ln N},
\end{eqnarray}
the correct black hole entropy will be certainly reproduced:
\begin{eqnarray}
S_{BH} \sim \frac{{\cal A}}{4G},
\end{eqnarray}
where ${\cal A}$ is the circumference of the horizon:
\begin{eqnarray}
{\cal A} =2\pi r_+ .
\end{eqnarray}
The last step of the entropy estimation must be said quite crude.
However, the same prescription also works in higher dimensional black
holes
 and reproduces the correct entropy form, ${\cal A}/4G$. Therefore it 
 may be possible that 
 the analysis indicates some  features of 
the exact quantum black hole physics, that is,
 the horizon might become a ``quantum 
 boundary" which is constructed from  
   area-unit elements of order of the planck scale
 on the horizon. To establish the exact estimation of the entropy,
 more information is clearly needed about the quantum gravity theory 
 and expected to be gained in the future work.

\ \\
\ \\
\ \\
{\bf \large Acknowledgement}
\ \\

I would like to thank J.Koga for helpful discussions.


\begin{thebibliography}{100}

\bibitem {c}
S.Carlip, Phys.Rev.{\bf D51},632,(1995).

\bibitem{bh}
J.D.Bekenstein, Phys.Rev.{\bf D7},2333,(1973).\\
S.W.Hawking, S.W.Hawking, Commun.Math.Phys.{\bf 43},199,(1975).

\bibitem{sv}
A.Strominger and C.Vafa, Phys.Lett.{\bf B379},99,(1996).

\bibitem{c1}
S.Carlip, Phys.Rev.Lett.{\bf 82},2828,(1999). 

\bibitem{v}
S.N.Solodukhin, Phys.Lett.{\bf B454},213,(1999).\\ 
F.-L.Lin and Y.-S.Wu, Phys.Lett.{\bf B453},222,(1999).\\
D.Navarro, J.Navarro-Salas, P.Navarro, Nucl.Phys.{\bf
B580},311,(2000).\\ 
R.Brustein, Phys.Rev.Lett.{\bf 86},576,(2001).\\ 
S.Das, A.Chosh and P.Mitra, Phys.Rev.{\bf D63},024023,(2001). \\
J.Jing and M.-L. Yan, Phys.Rev.{\bf D64},064015,(2001).

\bibitem{av}
M.-I.Park and J.Ho, Phys.Rev.Lett.{\bf 83},5595,(1999).\\
S.Carlip, Phys.Rev.Lett.{bf 83},5596,(1999).\\
M.-I. Park and J.H.Yee, Phys.Rev.{\bf D61},088501,(2000).\\
O.Dreyer, A.Ghosh and J.Wisniewski, \\
Class.Quant.Grav.{\bf 18},1929,(2001).\\ 
J.Koga, Phys.Rev.{\bf D64},124012,(2001).


\bibitem{hss}
M.Hotta, K.Sasaki and T.Sasaki, Class.Quant.Grav.{\bf 18},1823,(2001). 

\bibitem{wald}
J.Lee and R.M.Wald, J.Math.Phys.{\bf 31},725,(1990).\\
V.Iyer and R.M.Wald, Phys.Rev.{\bf D50},846,(1994);\\
Phys.Rev.{\bf D52},4430,(1995).


\bibitem{wz}
R.M.Wald and A.Zoupas, Phys.Rev.{\bf D61},084027,(2000).

\bibitem{wigner}
E.P.Wigner, Ann.Math.{\bf 40},149,(1939).


\end{thebibliography}
\end{document}